\documentclass{article}
\usepackage{graphicx}
\usepackage[round]{natbib}
\usepackage{epsfig}
\title{
Correlations between hurricane numbers and sea surface
temperature: why does the correlation disappear at landfall?
 }

\author{
Thomas Laepple (RMS and AWI)\\
Enrica Bellone (RMS)\\
Stephen Jewson (RMS)\footnote{\emph{Correspondence email}: \texttt{stephen.jewson@rms.com}}\\
Kechi Nzerem (RMS)\\
}

\begin{document}
\maketitle

\begin{abstract}
There is significant correlation between main development region
sea surface temperature and the number of hurricanes that form in the
Atlantic basin. The correlation between the same sea surface temperatures
and the number of \emph{landfalling} hurricanes is much lower, however. Why is this?
Do we need to consider complex physical hypotheses, or is there a simple
statistical explanation?
\end{abstract}

\section{Introduction}

Boreal summer sea surface temperatures (SSTs) in the tropical North Atlantic
correlate well with the number of hurricanes that form in the
Atlantic basin in the same year (see, for example, our own analysis in~\citet{e04a}, as well
as many previous analyses by other authors.).
This correlation forms the basis for a method for predicting future basin hurricane
numbers: first predict the sea surface temperature, then predict
the number of hurricanes given the sea surface temperature.
However, for many it is not the number of basin hurricanes that
is of most interest, but the number of hurricanes that make
landfall. And on a purely financial basis it is the number of
hurricanes making landfall in the US that is most important. It
thus makes sense to ask: how well do SSTs correlate with the
number of US landfalls? Much less well than they correlate with
the number of hurricanes in the basin, it turns out (see, for example, our own analysis
in~\citet{e04b}, as well as various previous analyses.). This is
perhaps a little surprising, since landfalling hurricanes are just
a subset of the hurricanes in the basin, and one might expect them
to inherit the same correlation with SST. We investigate what
could be causing this reduction in correlation, and, in
particular, consider a simple statistical model to see whether the
reduction in correlation should be expected on the basis of simple
statistical arguments. If it can, then there is little
motivation to consider more complex hypotheses for why the
correlation reduces, such as the possibility that the types of
hurricanes that make landfall are less affected by SSTs than other types of hurricanes.

\section{The effects of sea surface temperature on hurricane numbers}

In~\citet{e04a} and~\citet{e04b} we performed a thorough statistical
analysis of the relationship between main development region SSTs and the
number of hurricanes in the basin and at US landfall, respectively.
We considered a number of different periods of historical
data, and a number of different statistical models that one could use to relate the two.

The main overall conclusion: relating SST to basin hurricanes is much easier than relating SST to landfalling
hurricanes. Comparing~\citet{e04a} and~\citet{e04b}, this can be illustrated by consideration of:

\begin{itemize}

    \item \emph{linear correlations}: the linear correlations between SST and hurricane numbers are around 0.5 for the
    basin, and around 0.15 for landfalling hurricanes

    \item \emph{rank correlations}: the rank correlations between SST and hurricane numbers are around 0.5 for the
    basin, and around 0.2 for landfalling hurricanes

    \item \emph{standard errors on regression slopes}: when we fit linear regression lines between hurricane numbers
    and SST we find much larger (percentage) standard errors on the slopes of the lines between SST and landfalling
    numbers than we do between SST and basin numbers

    \item \emph{scatter plots}: the relationship between SST and hurricane numbers is very clear for basin hurricanes,
    and not at all clear for landfalling hurricanes

    \item \emph{regression-based predictive studies}: when we fit regression models between SST and hurricane numbers
    we find that we can make much better predictions of basin numbers than we can of landfalling numbers

\end{itemize}

These results are qualitatively robust to changes in the period of data used and to using intense hurricanes
only.

\section{Method}

One possible explanation for the reduction in the SST-hurricane number correlation at landfall is that this
is purely a statistical effect. We now investigate whether this is a reasonable hypothesis
by running Monte Carlo simulations to generate artificial basin and landfall data. We then look
at how the correlation with SST changes from basin to landfall in this artificial data set.

We first create 5000 realisations of 50 years of surrogate SST data, by
resampling the historical SST data for the period 1950-1999.
We then create 5000 realisations of 50 years of surrogate basin hurricane number data, based on these
SSTs, using the assumption that the number of hurricanes in a year is poisson distribution with mean
given by $6+5$x$SST$, which is a simple approximation to the SST-hurricane number relations fitted
in~\citet{e04a}.
Finally we create 5000 realisations of 50 years of surrogate landfall hurricane number data,
based on the simulated number of hurricanes in the basin in each year, and using a constant probability
that a basin hurricane will make landfall of $0.25$.

We then repeat the whole experiment using 135 years of surrogate SST data.

\section{Monte Carlo results}

\begin{figure}[!hb]
  \begin{center}
    \scalebox{0.7}{\includegraphics{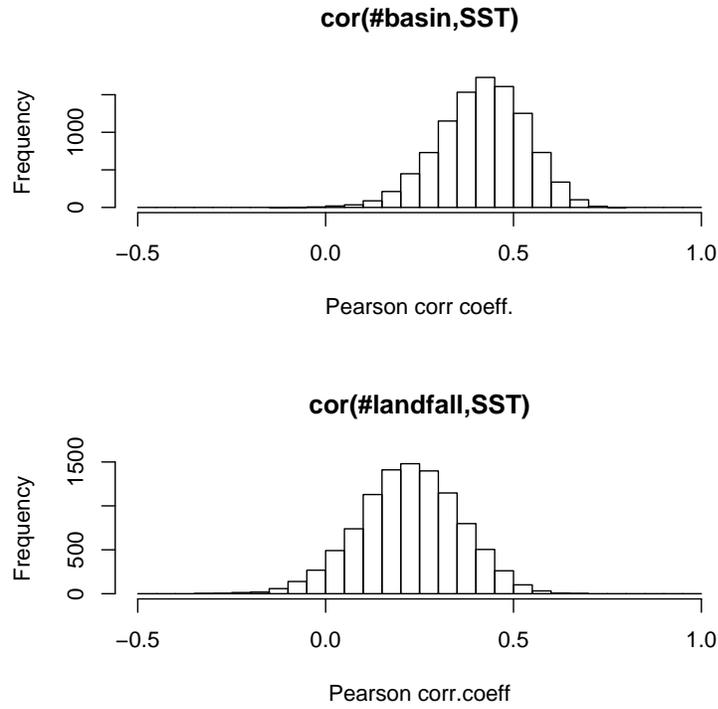}}
  \end{center}
    \caption{
The distribution of correlations between SST and the number of hurricanes in the basin
(upper panel) and at landfall (lower panel) based on 50 years of simulated historical data.
}
     \label{f01}
\end{figure}

We now present the results from our Monte Carlo tests with 50 years of data.
First, we consider the linear correlation between SST and hurricane numbers, in the basin and at landfall.
The top panel of figure~\ref{f01} shows a histogram of the linear correlations for SST and basin numbers, based on the
5000 realisations in our simulations (which give 5000 estimates for this correlation).
The mean of the distribution is 0.48, with some considerable spread.
The lower panel of figure~\ref{f01} shows a histogram of the linear correlations for SST and \emph{landfalling} hurricane
numbers, based on the same realisations.
The distribution has shifted significantly to the left: the mean of the distribution is now at 0.26, and again
there is considerable spread.
This illustrates immediately that we should expect a reduction in the correlation with SST
as a simple statistical consequence of moving
from basin to landfall.

\begin{figure}[!hb]
  \begin{center}
    \scalebox{0.7}{\includegraphics{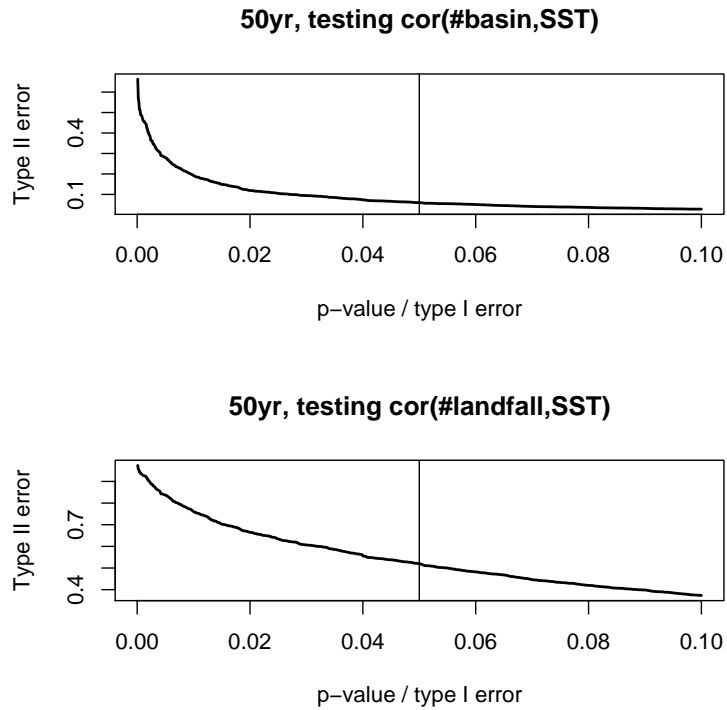}}
  \end{center}
    \caption{
The probability of rejecting the hypothesis that there is a correlation
between SST and hurricane numbers, given the p-value used, for basin hurricane numbers
(upper panel) and landfalling hurricane numbers (lower panel).
}
     \label{f03}
\end{figure}

We now ask the question: if we were to do tests of the significance of these correlation values, how frequently
would we conclude that the correlations are significant in the two cases? This is illustrated in figure~\ref{f03},
which shows the probability that we conclude there is \emph{not} a signficant relationship between SST and
hurricane numbers, versus the p-value used, for basin (top panel) and landfall (lower panel).
At p=0.05 we only conclude that there is no significant relationship 6\% of the time in the basin,
while at landfall we come to that conclusion 52\% of the time.

\begin{figure}[!hb]
  \begin{center}
    \scalebox{0.7}{\includegraphics{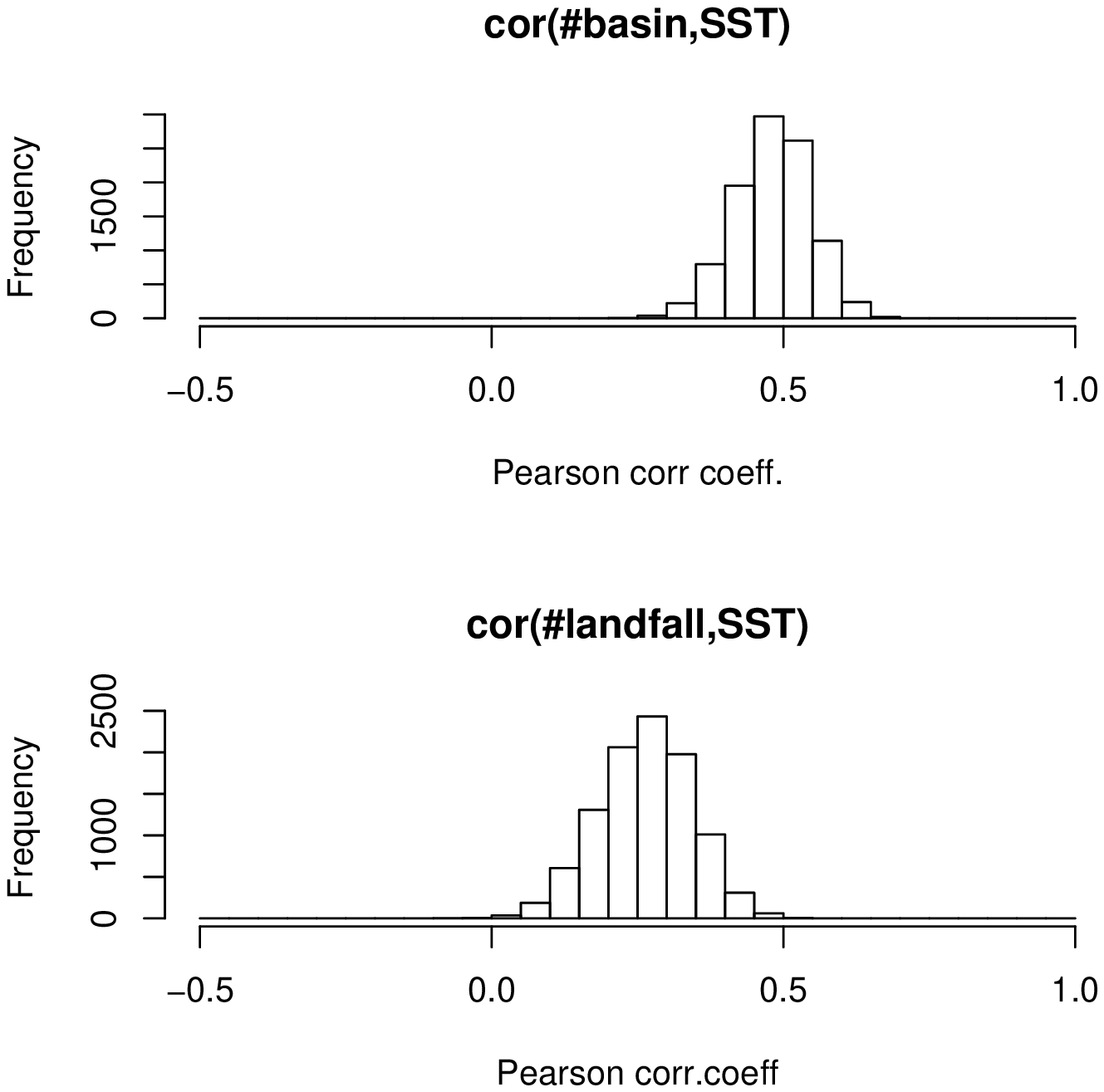}}
  \end{center}
    \caption{
The same as figure~\ref{f01}, but for 135 years of simulated historical data.
}
     \label{f02}
\end{figure}

\begin{figure}[!hb]
  \begin{center}
    \scalebox{0.7}{\includegraphics{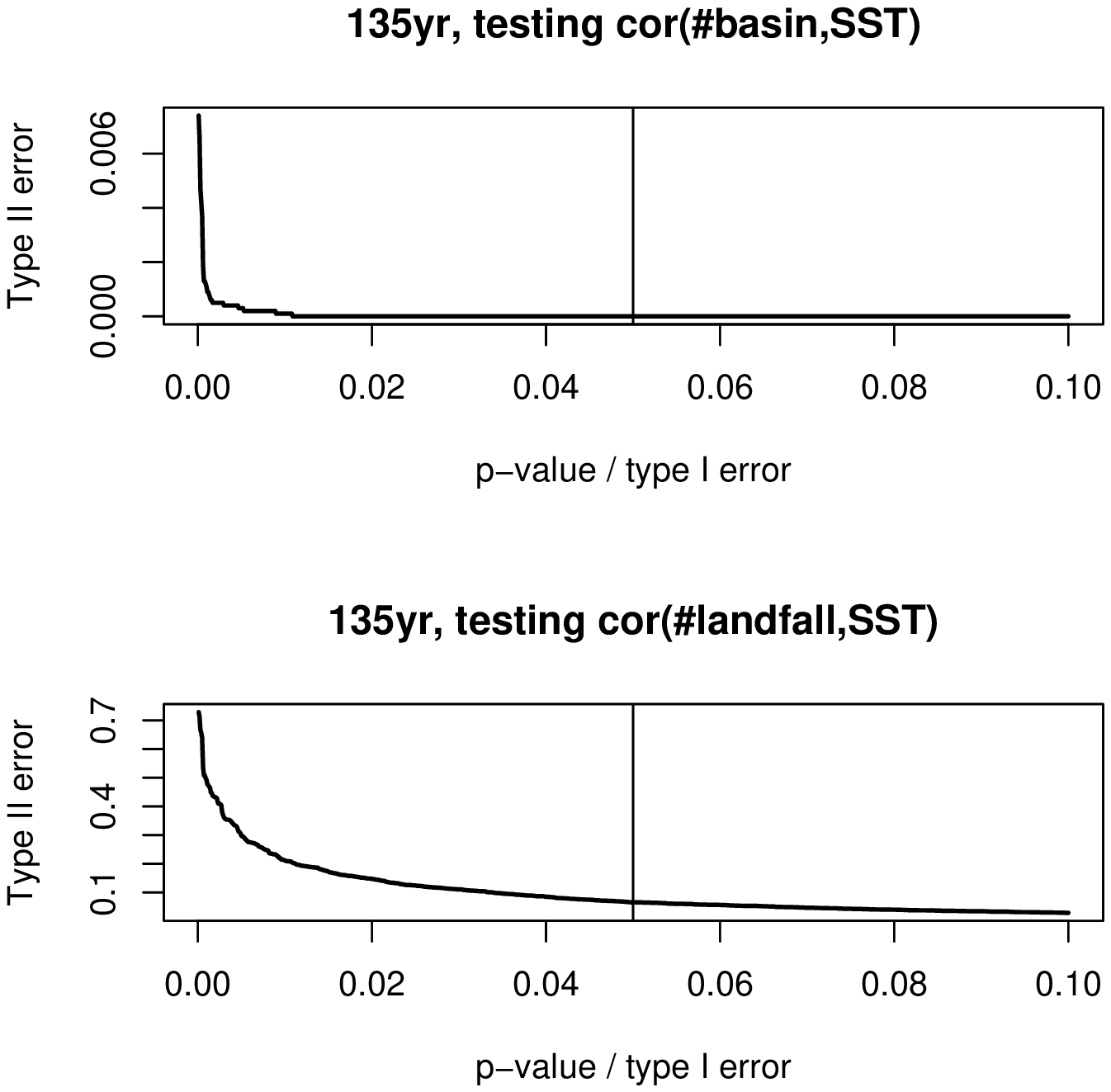}}
  \end{center}
    \caption{
The same as figure~\ref{f03}, but for 135 years of simulated historical data.
}
     \label{f04}
\end{figure}

We now show results for our Monte Carlo tests with 135 years of data.
Figure~\ref{f02} is the adapted version of figure~\ref{f01}, and figure~\ref{f04} is the adapted version
of figure~\ref{f03}. The correlations shown in figure~\ref{f02} have the same means as before, but the
distribution is now narrower, as a result of using more data in the estimation process.
The probabilities of concluding that there is no signficant relationship reduce considerably.
In the basin case, at p=0.05, the probability is less than 0.01\%: we are almost guaranteed to detect
the relationship if we have this much data. At landfall the probability of missing the relationship is 7\%.

\section{Conclusions}

Observed SSTs correlate at around 0.5 with Atlantic basin hurricane numbers, and at around 0.2 for landfalling
hurricane numbers. Why is there this big reduction at landfall? We have shown that one would expect
just such a big reduction \emph{purely on statistical grounds}: reducing the amount of data by reducing the numbers
of hurricanes per year reduces our ability to detect correlations with SST. This does not, of course,
prove that there aren't other things going on as well that might affect the correlation, but it would seem
likely that this is the principal effect.

\bibliography{arxiv}

\newpage

\end{document}